\documentclass[aps,prl,10pt,a4paper,twocolumn,footinbib,superscriptaddress,showkeys]{revtex4-2}
\usepackage[utf8]{inputenc}
\usepackage{fancyhdr}
\usepackage{amsmath}
\usepackage{amssymb}
\usepackage{amsthm}
\usepackage{graphicx}
\usepackage{dsfont}
\usepackage{makecell}
\usepackage{yfonts}
\usepackage{MnSymbol}
\usepackage{physics}
\usepackage{xcolor}
\usepackage{balance}
\usepackage{hyperref}

\newcommand{\Mod}[1]{\ (\mathrm{mod}\ #1)}

\begin{document}
\bibliographystyle{apsrev4-2}
\setcounter{secnumdepth}{2}

\title{Fast and robust cat state preparation utilizing higher-order nonlinearities in Rydberg ensembles}

\author{S. Zhao}
\affiliation{Institute for Quantum Materials and Technology, Karlsruhe Institute of Technology, 76344 Eggenstein-Leopoldshafen, Germany}
\affiliation{Freie Universit\"{a}t Berlin, Fachbereich Physik and Dahlem Center for Complex Quantum Systems, Arnimallee 14, 14195 Berlin, Germany}

\author{M. G. Krauss}
\affiliation{Freie Universit\"{a}t Berlin, Fachbereich Physik and Dahlem Center for Complex Quantum Systems, Arnimallee 14, 14195 Berlin, Germany}

\author{T. Bienaimé}
\affiliation{CESQ/ISIS (UMR 7006), CNRS and University of Strasbourg, 67000 Strasbourg, France}

\author{S. Whitlock}
\affiliation{CESQ/ISIS (UMR 7006), CNRS and University of Strasbourg, 67000 Strasbourg, France}

\author{C. P. Koch}
\affiliation{Freie Universit\"{a}t Berlin, Fachbereich Physik and Dahlem Center for Complex Quantum Systems, Arnimallee 14, 14195 Berlin, Germany}

\author{S. Qvarfort}
\affiliation{Nordita, KTH Royal Institute of Technology and Stockholm University, Hannes Alfv\'{e}ns v\"{a}g 12, SE-106 91 Stockholm, Sweden}
\affiliation{Department of Physics, Stockholm University, AlbaNova University Center, SE-106 91 Stockholm, Sweden}

\author{A. Metelmann}
\affiliation{Institute for Theory of Condensed Matter, Karlsruhe Institute of Technology, 76131 Karlsruhe, Germany}
\affiliation{Institute for Quantum Materials and Technology, Karlsruhe Institute of Technology, 76344 Eggenstein-Leopoldshafen, Germany}
\affiliation{CESQ/ISIS (UMR 7006), CNRS and University of Strasbourg, 67000 Strasbourg, France}
\affiliation{Freie Universit\"{a}t Berlin, Fachbereich Physik and Dahlem Center for Complex Quantum Systems, Arnimallee 14, 14195 Berlin, Germany}
 
\date{\today}
\begin{abstract}
In optical and solid-state architectures, low-order nonlinearities are commonly exploited for quantum state preparation due to their practical accessibility and controllability, while higher-order contributions are typically much weaker and treated as unwanted perturbations. Here, we alternatively show that detuned Rydberg ensembles provide a natural platform where higher-order Kerr nonlinearities can become comparable in strength to lower-orders near multiphoton resonances. We demonstrate that these nonlinearities can be harnessed as a resource for the rapid preparation of non-Gaussian states, with the coexistence of multiple Kerr orders substantially accelerating the evolution of an initial coherent state into Schr\"odinger cat states. Furthermore, by combining the nonlinear dynamics with a controllable linear drive, we can gain the full control over the evolution trajectory, thereby achieving cat-state generation directly from vacuum on timescales beyond the genuine Kerr evolution. Our results establish a paradigm in which naturally occurring higher-order nonlinearities are transformed from unwanted imperfections into a valuable resource for quantum state engineering.
\end{abstract}

\maketitle


\textit{Introduction.}--To enable advanced quantum technologies and realize complex quantum phenomena, the introduction of nonlinearity is essential. Here, nonlinearity refers to anharmonicity in the spectrum of a Hamiltonian or nonlinear equations of motion for operators. In general, nonlinearity may be provided by coupling bosonic modes to finite dimensional systems, or embedding them into a nonlinear medium~\cite{HaroldOxford06}. Among the various nonlinear interactions, Kerr-type nonlinearities play a central role due to their ability to induce nonlinear phase shifts that depend on arbitrary powers of the occupation number in the system, while preserving the population distribution and total energy. In particular, second-order Kerr interactions (quadratic in the occupation number) provide a versatile mechanism for universal control of bosonic systems~\cite{LloydPRL99, YuanPRA25, ErikssonNC24, RanganPRL04}, enabling the preparation of non-Gaussian quantum states such as Schrödinger cat states~\cite{PurinQI17, YurkePRL86, GrimmN20, KraussPRR23, TycNJP08, WiesJOB04, JoshiPLA00, HeNC23, BourassaPRA12, FrattiniAPL17, KhazaliPRA16, OmranS19}. These non-Gaussian states are an essential resource for quantum information processing~\cite{EisertPRL02, GiedkePRA02, MenicucciPRL06, AlbertQST19, GertlerN21, LescanneNP20a} and quantum enhanced sensing~\cite{LinnemannPRL16, YousefjaniPRA25, XuPRP19, ZwierzPRL10}.

In optical and solid-state platforms, second-order Kerr interactions are typically accompanied by higher-order Kerr nonlinearities, which are generally much weaker and hence often neglected or regarded as unwanted perturbations. Their influence becomes appreciable only in strongly driven systems, where they give rise to undesirable effects in higher-harmonic generation~\cite{GrynkoPRA17, WeerawarnePRL15} and saturation of parametric amplifiers~\cite{HillmannPRA22, RehakAPL14, LiuPRA20}. In contrast, interacting Rydberg ensembles provide an alternative platform where not only appreciable conventional Kerr nonlinearities emerage~\cite{KeatingPRL16, MorgadoAQS21, SaffmanRMP10, AdamsJPBAMOP19, vanBijnenPRL15}, but simultaneously also strong higher-order Kerr interactions~\cite{Henkel14, BaiOEO16, JachymskiPRL16, BieniasNJP16}. However, the simultaneous coexistence of multiple Kerr orders and their collective role in quantum state engineering remain largely unexplored. Existing perturbative treatment can accurately describe the full system dynamics~\cite{Henkel14}, but does not readily reveal the underlying hierarchy of Kerr nonlinearities. Understanding and harnessing this hierarchy is therefore essential for concretely exploiting higher-order Kerr nonlinearities as a useful resource for quantum state engineering.

In this work, we show that the unavoidable coexistence of multiple Kerr orders in Rydberg ensembles can be harnessed as a powerful resource, as their collective induced dynamics substantially accelerate cat state preparation~\cite{YurkePRL86}. While the acceleration typically relies on a fine-tuning of the Kerr coefficients, we demonstrate that an optimally controlled linear drive fully exploits this advantage without requiring such fine-tuning, enabling substantially faster cat-state preparation directly from vacuum. Beyond the Rydberg framework, our work reveals a general strategy to harness naturally occurring higher-order nonlinearities as a resource for accelerated quantum state engineering.

\begin{figure}
    \includegraphics[width=0.95\columnwidth]{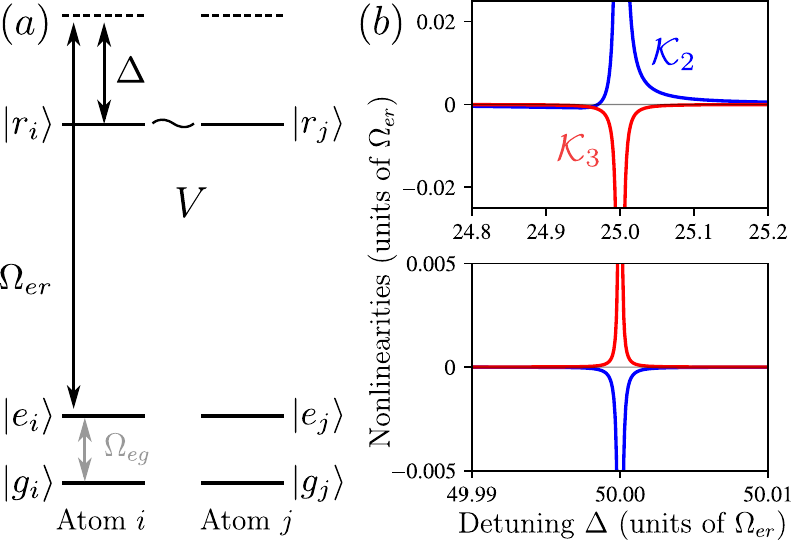}
    \caption[Rydberg plots]{(a) Energy diagram of two three-level atoms interacting via the Rydberg-Rydberg interaction $V$. $|g\rangle$, $|e\rangle$ and $|r\rangle$ denote ground, excited and Rydberg states. The laser with Rabi frequency $\Omega_\mathrm{er}$ (or $\Omega_\mathrm{eg}$ in gray) drives the transition $e\leftrightarrow r$ off-resonantly with a detuning $\Delta$ (or $g\leftrightarrow e$ resonantly). (b) Coefficients of the second- and third-order Kerr interactions $\mathcal{K}_2$ (in red) and $\mathcal{K}_3$ (in blue), plotted against $\Delta$ for a fixed $V=50\Omega_\mathrm{er}$. Strong nonlinearities occur in the vicinity of resonances $V=\frac{2\Delta}{m}$ for $m=0, 1,2,...$ between excited state subspace and $m$-Rydberg subspaces.}
    \label{fig_rydberg}
\end{figure}

\textit{Model.}--We consider a collection of Rydberg atoms with a level structure as illustrated in Fig. \ref{fig_rydberg}(a). The corresponding Hamiltonian yields \begin{align}
    \hat{H}_{\mathrm{full}}&=\sum_{i=1}^N\bigg(\frac{\Omega_\mathrm{er}}{2}(\hat{\sigma}_i^{re}+\hat{\sigma}_i^{er})-\Delta\hat{\sigma}_i^{rr}+V\sum_{j>i}^N \hat{\sigma}_i^{rr}\hat{\sigma}_j^{rr}\bigg), \label{eq_rydberg_full_hamiltonian}
\end{align}
where $\Omega_\mathrm{er}$ is the Rabi frequency of a laser that drives the transition between excited and Rydberg states off-resonantly by a detuning $\Delta$, and $\hat{\sigma}_i^{xy}=|x\rangle\langle y|$ denotes the jump operator acting on the $i$-th atom. Here, the Rydberg interaction has been approximated by an ensemble-averaged coupling $V$ (see End Matter). This approximation captures the overall collective nonlinearity while neglecting geometry-dependent fluctuations, allowing the dynamics to be described within the symmetric Dicke manifolds coupled by the collective laser drive $\Omega_\mathrm{er}$. To obtain an effective Hamiltonian acting on the reduced $\left\{|g\rangle,|e\rangle\right\}$ subspace, we utilize the projector method~\cite{OlmosPRA09} to adiabatically eliminate the Rydberg excitations from the full system. For simplicity, Eq.~\eqref{eq_rydberg_full_hamiltonian} excludes the laser drive $\Omega_{eg}$ (gray arrow in Fig.~\ref{fig_rydberg}(a)), as it does not affect the adiabatic elimination of the Rydberg states. It will instead be incorporated later as a controllable linear drive for optimal control. In the limit of large detunings from resonances between the excited and Rydberg subspaces, i.e., $\Delta-mV/2\gg \Omega_\mathrm{er}$ for $m=0,1,2,\dots$, the resulting Hamiltonian can be written as a power series expansion (see End Matter)\begin{align}
    \hat{H}_{\mathrm{eff}}&=\sum_{n=2}^N\mathcal{K}_n\hat{N}_{e}^n \label{eq_rydberg_effective_hamiltonian},
\end{align}
where the eigenstates correspond to states with definite numbers of collective excitations in the ensemble, or symmetrized Dicke states~\cite{KeatingPRL16, RobertPRA26}, which are analogous to Fock states in their bosonic representation. Although previous perturbative treatments of the full Hamiltonian provide higher accuracy, they quickly become analytically intractable at higher orders. Here, we instead developed an adiabatic elimination approach within the lower levels of the hierarchy of quantum many-body approximations~\cite{PaulischEPJP14}, enabling systematic access 
to arbitrarily high-order nonlinearities.

In Eq.~\eqref{eq_rydberg_effective_hamiltonian}, we moved to a rotating frame about the linear term $\mathcal{K}_1\hat{N}_e=\mathcal{K}_1\sum_{i=1}^N\hat{\sigma}_i^{ee}$, and the nonlinear coefficients can be expressed as $\mathcal{K}_n=\sum_{m=0}^{N-1}s(m+1,n)f(m)$, with $s(m,n)$ being the (signed) Stirling number of first kind and \begin{align} 
    f(m)&=\begin{cases}
    \frac{\Omega_\mathrm{er}^2}{4\Delta} &\hspace{-0.27cm}m=0,\\
    \frac{1}{(m-1)!}\frac{V\Omega_\mathrm{er}^{2+2m}}{2^{3+2m}\Delta(\Delta-\frac{m}{2}V)\prod_{k=1}^m(\Delta-\frac{k-1}{2}V)^2}\quad&\text{else}.\hspace*{-0.3cm}
    \end{cases}
\end{align}
\begin{figure}
    \includegraphics[width=0.9\linewidth]{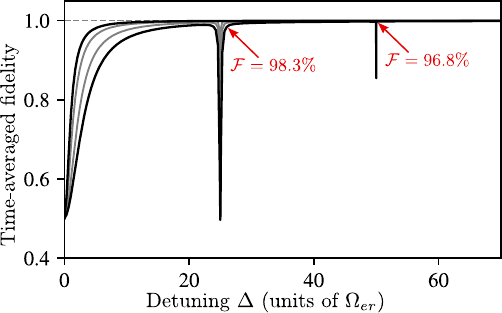}
    \caption{Time-averaged fidelity between an initial excited state evolving under the full and effective Hamiltonians for $N=1,2,5,10$ atoms with $V=50\Omega_\mathrm{er}$. An ideal fidelity of unity indicates perfect agreement between the full and effective models. The spikes at $\Delta=mV/2$ for $m=0,1,2,\dots$ correspond to the $m+1$-photon resonance between the excited subspace and $m+1$-Rydberg subspace. The fidelity rapidly decreases at these resonance regions, indicating the breaking down of our effective model. The red arrows at $\Delta=26\Omega_\mathrm{er}$ and $\Delta=50.0086\Omega_\mathrm{er}$ indicate regions where strong nonlinearities can be achieved while maintaining high accuracy of the effective model.}
    \label{fig_full_eff_comparison}
\end{figure}

Eq.~\eqref{eq_rydberg_effective_hamiltonian} consists of multiple orders of Kerr nonlinearities in the excitation number operator $\hat{N}_e$, and highlights their unavoidable coexistence in realistic scenarios. These Kerr nonlinearities can diverge due to singularities arising from a vanishing denominator $\Delta-mV/2=0$ of $f(m)$ for $m=0,1,2,\dots$, where the resonance at $\Delta=mV/2$ involves the $m+1$-Rydberg excitation manifold, and therefore corresponds to an $m+1$-photon process~\cite{GarttnerPRL14, ShorePRA81}. Fig.\ref{fig_rydberg}(b) shows the divergence of $\mathcal{K}_2$ and $\mathcal{K}_3$ at two- and three-photon resonances ($m=1,2$). These singularities signal the breakdown of our effective Hamiltonian, yet they also mark the regime where strong higher-order Kerr nonlinearities can be achieved. We therefore prefer to operate in the vicinity of these singular regions for strong Kerr nonlinearities, while the effective Hamiltonian Eq.~\eqref{eq_rydberg_effective_hamiltonian} remains sufficiently accurate.

\begin{figure}
    \includegraphics[width=0.9\linewidth]{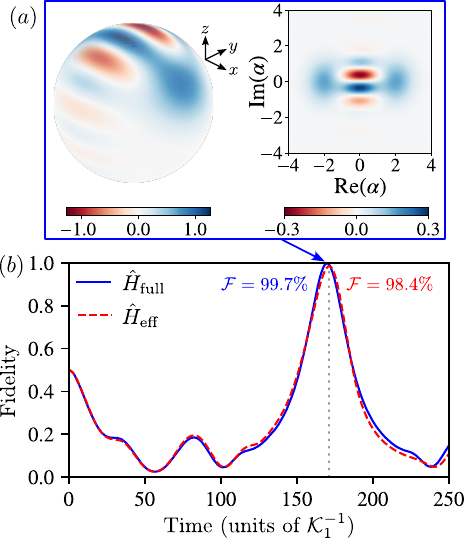}
    \caption{(a) Evolution of an initial (atomic) coherent state $|N,0.5\rangle\sim(|g\rangle+0.5|e\rangle)^{\otimes N}$ under full and effective Hamiltonians into a cat state at time $t_c\sim 170 \mathcal{K}_1^{-1}$ (dotted vertical line), using the parameters $N=10$, $\Omega_\mathrm{er}=1$, $\Delta=26\Omega_\mathrm{er}$, $V=50\Omega_\mathrm{er}$ (marked by the first red arrow in Fig.~\ref{fig_full_eff_comparison}). (b) The generated cat state is plotted on both the Bloch sphere and complex plane.}
    \label{fig_kerr_cat_evol}
\end{figure}

\textit{Higher-order Kerr.}--We find that higher-order Kerr nonlinearities can substantially accelerate the Kerr evolution used to generate Schr\"odinger cat states. We first note that similar to how an initial bosonic coherent state can evolve into a cat state under a single-order Kerr interaction at a particular time $t_c$~\cite{YurkePRL86}, this can be similarly applied to atomic cat states \begin{align}
    e^{\pm i\mathcal{K}_mt_c\hat{N}_e^m}|N,\xi\rangle\sim |N,\xi\rangle\pm i|N,-\xi\rangle, \label{eq_yurke_cat_evol}
\end{align}
where $|N,\xi\rangle\sim (|g\rangle+\xi|e\rangle)^{\otimes N}\sim\sum_{n=0}^N \sqrt{\binom{N}{n}}\ \xi^n |n,N-n\rangle$ is the atomic coherent state, $\binom{N}{n}$ is the binomial coefficient, and $|n,N-n\rangle\equiv \{|e\rangle^{\otimes n}\otimes |g\rangle^{\otimes (N-n)}\}_\mathrm{sym}$ denotes the symmetrized atomic Dicke state with $n$ and $N-n$ atoms in excited and ground states respectively, which can be viewed as the atomic analogue of the bosonic Fock state $|n\rangle$. In particular, in the limit of large $N$, the underlying binomial statistics of the atomic coherent state agrees with the Poisson statistics of the bosonic coherent state, and thus the two distinct types of systems become mathematically equivalent~\cite{CahenBZAG03, RicciMFM86}. For simplicity, in the following, we will keep the bosonic description of the Rydberg ensemble, with which Eq.~\eqref{eq_rydberg_effective_hamiltonian} becomes
\begin{align}
    \hat{H}&=\sum_{j=2}^m\mathcal{K}_j(\hat{a}^\dag\hat{a})^j,
    \label{eq_rydberg_bosonic_kerr}
\end{align}
where $\hat{a}^\dag\hat{a}\equiv \hat{N}_e$. Equation~\eqref{eq_rydberg_bosonic_kerr} highlights the coexistence of multiple orders of Kerr nonlinearities in realistic platforms, with each contributing a distinct nonlinear phase to the Fock state components of the system. For particular combinations of Kerr coefficients, the resulting nonlinear phases can collectively reproduce the phase structure of a cat state. Consequently, despite the presence of multiple higher-order Kerr terms, cat states can still be generated in suitable parameter regimes~\cite{supp_mat}. For example, near the two-photon resonance indicated by the first red arrow in Fig.~\ref{fig_full_eff_comparison}, our effective model predicts a Kerr configuration of $\mathcal{K}_2=0.01\mathcal{K}_1$ and $\mathcal{K}_3=0.01\mathcal{K}_2$, which yields a cat formation time $t_c\sim 170\mathcal{K}_1^{-1}$ as shown in Fig.~\ref{fig_kerr_cat_evol}(b). The resulting cat state is plotted on both a Bloch sphere and a complex plane in Fig.~\ref{fig_kerr_cat_evol}(a), demonstrating the equivalence between atomic and bosonic descriptions in the limit of large $N$.

Moreover, when higher-order Kerr nonlinearities are sufficiently strong, their phase contributions become more significant. For instance, the second red arrow in Fig.~\ref{fig_full_eff_comparison} indicates a region where the third-order Kerr is as strong as the second-order $\mathcal{K}_3=-0.53\mathcal{K}_2$. This provides more substantial degrees of freedom in tuning the Kerr coefficients to match the phase structure required for a cat state. Intuitively, since the acquired nonlinear phase scales as $\mathcal{K}_{\bullet}t_c$, these significant extra degrees of freedom can cooperate and enable faster cat-state preparation. However, once the induced phases exceed the $2\pi$ periodicity, further increasing in nonlinearities becomes ineffective, leading to an optimal minimum preparation time~\cite{supp_mat} \begin{align}
    t_{c,\mathrm{min}}&=\begin{cases}
        \frac{\pi}{2\mathcal{K}_2} \hspace{0.5cm}&\text{for $m=2$},\\
        \frac{2\pi}{\mathcal{K}_mm!} &\text{for $m\geq 3$}. \label{eq_tc_def}
    \end{cases}
\end{align}

\begin{figure}
    \includegraphics[width=0.9\columnwidth]{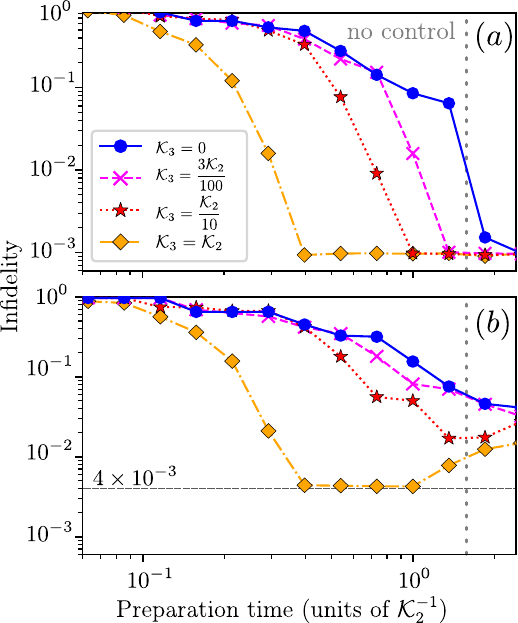}
    \caption{Infidelity between the target cat state and those prepared using optimal control, against the pulse duration (or cat preparation times). All markers correspond to the lowest infidelity obtained from a series of optimizations that are differed by a guess pulse. The same optimizations are run for both (a) closed system, and (b) open system with single-photon loss and dephasing $\kappa_{1\mathrm{ph}}=\kappa_\phi=3\times10^{-3}\mathcal{K}_2$. The vertical dotted line marks the shortest preparation time without employing control.}
    \label{fig:oct}
\end{figure}
\textit{Quantum control.}--While the accelerated Kerr evolution enables significantly faster cat-state preparation, its performance relies on tuning to a particular configuration of Kerr coefficients, and an initially prepared coherent state. However, finding suitable Rydberg parameters that realize the required configuration while maintaining sufficiently strong nonlinearities can be challenging. For example, the strong third-order Kerr interaction $\mathcal{K}_3=-0.53\mathcal{K}_2$ marked by the second arrow in Fig.~\ref{fig_full_eff_comparison} is deviated from the ideal value $\mathcal{K}_3=-2/3\mathcal{K}_2\approx-0.67\mathcal{K}_2$. Moreover, preparing the initial coherent state typically requires an additional linear drive, which will ruin the desired Kerr evolution. Although a sharply peaked pulse can be used to rapidly initialize the coherent state without affecting the subsequent Kerr evolution, its requirement already implies access to certain level of control over drive. This in turn naturally leads to the framework of quantum optimal control, where the potential of a controllable linear drive can be fully harvested throughout the Kerr evolution~\cite{GlaserEPJD15, KochEQT22}. In the following, we show that adding a controllable linear drive removes the need for fine-tuned Kerr configurations and enables the cat state preparation directly from vacuum. As a result, higher-order nonlinearities can be fully exploited to accelerate state preparation, with stronger nonlinearities leading to faster protocols.

To demonstrate this, we consider a linear drive \begin{align}
  \hat{H}_{\mathrm{dr}}&=\varepsilon(t)(\hat{a}+\hat{a}^\dagger),
  \label{eq:lineardrive}
\end{align}
where $\varepsilon(t)\in\mathds{R}$ is the pulse amplitude, which is the bosonic analogue of an atomic laser that drives the transition between ground and excited states in the limit of large $N$~\cite{CahenBZAG03, RicciMFM86} (gray arrow in Fig.~\ref{fig_rydberg}). We note that unlike the two-photon drive employed in Ref.~\cite{CochranePRA99, GrimmN20, PurinQI17}, a single linear drive is already sufficient to achieve full controllability over the system~\cite{LloydPRL99, KeatingPRL16}. This full controllability removes the need for a specific initial state in the cat state preparation, i.e., an initial vacuum is already enough. For our optimal control strategy, we employ Krotov's method~\cite{Krotov95,KonnovARC99,GoerzSP19} to minimize the infidelity $1-\left|\langle\psi_\mathrm{cat}|\psi(T)\rangle\right|^2$ between the optimized state $|\psi(T)\rangle$ prepared at final time $T$ and the target cat state $|\psi_\mathrm{cat}\rangle$ Eq.~\eqref{eq_yurke_cat_evol}. By repeating the optimization for several $T$, we estimate the shortest possible preparation time of a cat state, similar to Ref.~\cite{CanevaPRL09}. For an experimentally relevant description, we set the upper bound of the pulse amplitude to $|\varepsilon(t)|\leq 30\mathcal{K}_2$.

Fig.~\ref{fig:oct}(a) shows the optimization results for different values of $\mathcal{K}_3$. For each curve, the infidelity monotonously decreases towards larger preparation times and ultimately reaches a plateau at $10^{-3}$, where we stop the optimization and consider the result to be converged. Thus, the shortest possible preparation time is marked by the transition to the plateau. Comparing the curves in Fig.~\ref{fig:oct}(a) we find a clear shift towards shorter preparation times for larger values of $\mathcal{K}_3$. Moreover, when quantum optimal control is employed, the Kerr coefficients $\mathcal{K}_3=-0.53\mathcal{K}_2$ marked by the second red arrow in Fig.~\ref{fig_full_eff_comparison} yield a cat state preparation time close to that of the orange curve ($\mathcal{K}_3=\mathcal{K}_2$) in Fig.~\ref{fig:oct}(a). In contrast, without optimal control, the genuine Kerr evolution method is bounded by a minimum cat-generation time $t_{c,\mathrm{min}}=\pi/(2\mathcal{K}_2)$, indicated by the vertical dotted line. Fig.\ref{fig:oct}(b) presents the same results, but including single-photon loss and dephasing $\kappa_{1\mathrm{ph}}=\kappa_\phi= 3\times 10^{-3} \mathcal{K}_2$. Although the final infidelities are larger due to the presence of noise, a strong $\mathcal{K}_3$ still yields a clear improvement compared to $\mathcal{K}_3=0$. The orange curve exhibits a plateau within $0.4\mathcal{K}_2\leq T\leq 1.0\mathcal{K}_2$. This behavior can be explained by the control strategy, which maintains the system in the ground state for an extended duration before initiating the preparation process towards the very end of the protocol. Due to the inherent protection of the ground state, noise impacts the fidelity only during the fast cat preparation phase. These results indicate that strong higher-order Kerr nonlinearities can help mitigate the detrimental effects of decoherence by speeding up the preparation. We further note that, following the robust preparation of cat states, optimal squeezing can be employed to enhance their life-time in a noisy environment~\cite{supp_mat}.

\textit{Conclusion.}--To conclude, we derived an effective Hamiltonian using adiabatic elimination to study the coexistence of multiple orders of strong Kerr nonlinearities in Rydberg ensembles. We showed that suitable configurations of Kerr coefficients can substantially accelerate the evolution of an initial coherent state into cat states. Furthermore, by introducing a controllable linear (laser) drive, the needs for specific configurations of Kerr coefficients and the initial coherent state are removed. This enables the preparation of a cat state directly from the vacuum at a rather shorter time. Our results establish a paradigm in which naturally occurring higher-order nonlinearities are harnessed as a valuable resource, rather than treated as a detrimental effect.


We acknowledge fruitful discussions with F. Gago Encinas and L. Orr.
AM, MGK and CPK acknowledge financial support from the Deutsche Forschungsgemeinschaft (DFG), Project No. 277101999, CRC 183 (project C05).
TB acknowledges financial support from the Agence Nationale de la Recherche (ANR) through the SIQ-FLight project (Grant No. ANR-22-CE30-0043).
SW acknowledges support under the Investments of the Future Program with the reference ANR-21-ESRE-0032 and the “EuRyQa” project grant agreement number 10107014 and support from the Institut Universitaire de France (IUF).
SQ is funded in part by the Wallenberg Initiative on Networks and Quantum Information (WINQ) and in part by the Marie Skłodowska--Curie Action IF programme “Nonlinear optomechanics for verification, utility, and sensing” (NOVUS) -- Grant-Number 101027183. Nordita is partially supported by Nordforsk.
AM acknowledges funding by the Deutsche Forschungsgemeinschaft through the Emmy Noether program (Grant No.~ME 4863/1-1).

\balance
%


\onecolumngrid
\section*{End Matter}
\twocolumngrid
\renewcommand{\theequation}{E.\arabic{equation}}
\renewcommand{\thefigure}{E.\arabic{figure}}
\setcounter{equation}{0}
\setcounter{figure}{0}
\textit{Appendix: Effective Hamiltonian.}--In the following, we use adiabatic elimination~\cite{OlmosPRA09} to derive the effective Kerr Hamiltonian Eq.~\eqref{eq_rydberg_effective_hamiltonian}. Physically, each order of Kerr nonlinearities can be understood as arising from virtual excursions into subspaces containing a fixed number of Rydberg excitations. Eliminating these virtual processes induces energy shifts in the excited subspace (analogous to the Lamb shift). Since the $n$-Rydberg excitations can be distributed among any subset of the $N_e$ excited atoms, the resulting energy shift acquires a combinatorial factor $\binom{N_e}{n}\sim N_e^n$ that corresponds to the $n$-th-order Kerr shift.

We start from the full microscopic Hamiltonian of a Rydberg ensemble, whose atomic level structure is illustrated in Fig. \ref{fig_rydberg}(a) \begin{align}
    \hat{H}&=\sum_{i=1}^N\bigg(\frac{\Omega_\mathrm{er}}{2}(\hat{\sigma}_i^{re}+\hat{\sigma}_i^{er})-\Delta\hat{\sigma}_i^{rr}+\sum_{j>i}^NV_{ij}\hat{\sigma}_i^{rr}\hat{\sigma}_j^{rr}\bigg).\label{eq_rydberg_full_hamiltonian_Vij}
\end{align}
where $\Omega_\mathrm{er}$ is the Rabi frequency of the laser driving the transition between excited and Rydberg states, detuned by $\Delta$, $\hat{\sigma}_i^{xy}=|x\rangle\langle y|$ is the jump operator on the $i$-th atom. Motivated by the collective nature of the dynamics, in which the laser drive $\Omega_{\mathrm{er}}$ couples exclusively to the symmetric Dicke manifold, we approximate the pairwise Rydberg interactions $V_{ij}$ by an effective ensemble-averaged interaction $V=\sum_{i=1}^N\sum_{j>i}^NV_{ij}/(N(N-1))$. Consequently, the relevant states are not configurations with specific atoms excited to the Rydberg state, but rather the symmetrized Dicke states $\{|r\rangle^{\otimes N_r}\otimes|e\rangle^{N-N_r}\}_\mathrm{sym}\equiv|N_r,N_e\rangle$ that contains all permutations of the $N_r$ Rydberg states in the ensemble. To illustrate this explicitly, we first rewrite the Rydberg interaction term in Eq.~\eqref{eq_rydberg_full_hamiltonian_Vij} as
\begin{align}
    \sum_{i=1}^N\sum_{j>i}^N V_{ij}\hat{\sigma}_i^{rr}\hat{\sigma}_j^{rr}=\frac{\sum_{i=1}^N\sum_{j>i}^N V_{ij}\hat{\sigma}_i^{rr}\hat{\sigma}_j^{rr}}{\sum_{i=1}^N\sum_{j>i}^N \hat{\sigma}_i^{rr}\hat{\sigma}_j^{rr}}\sum_{i=1}^N\sum_{j>i}^N \hat{\sigma}_i^{rr}\hat{\sigma}_j^{rr}.
\end{align}
The expectation value of the prefactor with respect to the subspaces with $N_r\ge2$ evaluates to
\begin{align}
    \langle N_r,N_e|\frac{\sum_{i=1}^N\sum_{j>i}^N V_{ij}\hat{\sigma}_i^{rr}\hat{\sigma}_j^{rr}}{\sum_{i=1}^N\sum_{j>i}^N \hat{\sigma}_i^{rr}\hat{\sigma}_j^{rr}}|N_r,N_e\rangle=\frac{\sum_{i=1}^{N}\sum_{j>i}^{N}V_{ij}}{N(N-1)}.
    \label{eq_ensemble_average_V}
\end{align}
The key feature of Eq.~\eqref{eq_ensemble_average_V} is that the ensemble-averaged Rydberg interaction is identical across all symmetric Dicke manifolds, i.e., it is independent of the number of Rydberg excitations $N_r$. This naturally motivates the introduction of the ensemble-averaged interaction $V$, which captures the collective interaction energy while neglecting geometry-dependent fluctuations. The full macroscopic Hamiltonian is defined in Eq.~\eqref{eq_rydberg_full_hamiltonian}.

Based on Eq.~\eqref{eq_rydberg_full_hamiltonian}, our adiabatic elimination starts by dividing the full Hilbert space into subspaces containing a fixed number of Rydberg excitations. For this, we define the corresponding projectors as
\begin{align}
    \hat{P}  = \prod_{i = 1}^N 
    |e_{i} \rangle \langle e_{i} |,\ \  \hat{Q}_n =\frac{1}{n!}\sum_{\{i_n\neq\}}^N|r_{\{i_n\}}\rangle\langle r_{\{i_n\}}|, \label{Projector_def}
\end{align}
where $\{i_n\neq\}$ represents the summation over indices $i_1,\dots,i_n$ with $i_1\neq\dots\neq i_n$, $|r_{\{i_n\}}\rangle$ denotes the state with atoms labelled by $\{i_n\}$ being in Rydberg state, and the rest in excited state. Now, we make the Ansatz
\begin{align}
    \hat{H}_\mathrm{eff}=\sum_{n=0}^{N}\hat{H}_\mathrm{eff}^{(n)}.\label{eq_Heff_ansatz}
\end{align}
Here, $\hat{H}_\mathrm{eff}^{(n)}$ denotes the pure Rydberg-mediated $n$-body interaction, i.e., the additional interaction (or nonlinear response) arising from virtual transitions from $(n-1)$- to $n$-Rydberg subspaces, where the system propagates under the influence of Rydberg interaction. Starting from the $0$-Rydberg manifold $\hat{P}$, the hierarchy of many-body interactions can therefore be constructed recursively by comparing the Rydberg-mediated interaction increment of virtual transitions between adjacent subspaces with $n=1,2,3,\ldots$. For example, the transition $\hat Q_1\rightarrow\hat Q_2$ is allowed in the absence of Rydberg interactions $V=0$, but becomes increasingly suppressed for finite $V$, and is completely forbidden in the Rydberg-blockade limit $V\rightarrow\infty$. The resulting modification of the virtual transition relative to the non-interacting case induces a Rydberg-mediated nonlinear shift, which manifests as the effective two-body nonlinearity, i.e., the second-order Kerr interaction in Rydberg ensembles~\cite{KeatingPRL16,BalewskiNJP14}. Extending this construction to higher virtual transitions $\hat{Q}_n\rightarrow\hat{Q}_{n+1}$ naturally yields higher-order Kerr nonlinearities.

For $n=0$, where the system allows no Rydberg excitations, its corresponding Schr\"odinger equation is simply \begin{align}
    d\frac{d}{dt}(\hat{P}|\psi\rangle)&=\hat{P}\hat{H}\hat{P}(\hat{P}|\psi\rangle)=0,
\end{align}
which gives $\hat{H}_\mathrm{eff}^{(0)}=\hat{P}\hat{H}\hat{P}=0$. Whereas for $n=1$, where a maximum of one Rydberg excitation is allowed, the Schr\"odinger equation reads\begin{align}
    i\frac{d}{dt}\begin{pmatrix}
        \hat{P}|\psi\rangle\\
        \hat{Q}_1|\psi\rangle
    \end{pmatrix}&=\begin{pmatrix}
            \hat{P}\hat{H}\hat{P} & \hat{P}\hat{H}\hat{Q}_1 \\
            \hat{Q}_1\hat{H}\hat{P} & \hat{Q}_1\hat{H}\hat{Q}_1 \\
\end{pmatrix}\begin{pmatrix}
        \hat{P}|\psi\rangle\\
        \hat{Q}_1|\psi\rangle
    \end{pmatrix}, \label{eq_sch_Q1}
\end{align}
or equivalently \begin{align}
    \nonumber i\frac{d}{dt}\hat{P}|\psi\rangle&=\hat{P}\hat{H}\hat{P}\hat{P}|\psi\rangle+\hat{P}\hat{H}\hat{Q}_1\hat{Q}_1|\psi\rangle,\\
    i\frac{d}{dt}\hat{Q}_1|\psi\rangle&=\hat{Q}_1\hat{H}\hat{P}\hat{P}|\psi\rangle+\hat{Q}_1\hat{H}\hat{Q}_1\hat{Q}_1|\psi\rangle. \label{eq_sch_01}
\end{align}
Since in the large-detuning limit $|\Delta|\gg\Omega_\mathrm{er}$, transitions between the excited $\hat{P}$ and Rydberg $\hat{Q}_n$ subspaces are suppressed, the Rydberg subspace contribution to the excited dynamics becomes quasi-static, i.e., $\frac{d}{dt}\hat{Q}_n|\psi\rangle=0$ for $n=1,2,\dots$. Applying this condition (for $n=1$) back to the first line of Eq.~\eqref{eq_sch_01}, we obtain \begin{align}
    \hat{Q}_1|\psi\rangle=\left(-(\hat{Q}_1\hat{H}\hat{Q}_1)^{-1}\hat{Q}_1\hat{H}\hat{P}\right)\hat{P}|\psi\rangle,
\end{align}
with which the second line becomes \begin{align}
    i\frac{d}{dt}\hat{P}|\psi\rangle&=\left(\hat{H}_\mathrm{eff}^{(0)}-\hat{P}\hat{H}\hat{Q}_1(\hat{Q}_1\hat{H}\hat{Q}_1)^{-1}\hat{Q}_1\hat{H}\hat{P}\right)\hat{P}|\psi\rangle. \label{eq_adia_proj}
\end{align}
Here, the terms \begin{align}
    \hat{Q}_n\hat{H}\hat{Q}_{n+1}&=\frac{\Omega_\mathrm{er}}{2n!}\sum_{\{i_{n+1}\neq\}}\hat{\sigma}_{i_1}^{rr}\dots\hat{\sigma}_{i_n}^{rr}\ \hat{\sigma}_{i_{n+1}}^{er},
\end{align}
denote the transitions between adjacent Rydberg subspaces ($\hat{Q}_0\equiv\hat{P}$) for $n=0,1,2,\dots$, and
\begin{align}
    (\hat{Q}_n\hat{H}\hat{Q}_n)^{-1}&=\frac{-\sum_{\{i_n\neq\}}^N\hat{\sigma}_{i_1}^{rr}\dots\hat{\sigma}_{i_n}^{rr}}{n!(n\Delta-n(n-1)\frac{V}{2})},
\end{align}
denotes the propagator within the $n$-Rydberg subspace for $n=1,2,3,\dots$. Eq.~\eqref{eq_adia_proj} can be treated as the effective Schr\"odinger equation over the excited subspace, with the effective Hamiltonian
\begin{align}
    \hat{H}_\mathrm{eff}^{(1)}&=-\hat{P}\hat{H}\hat{Q}_1(\hat{Q}_1\hat{H}\hat{Q}_1)^{-1}\hat{Q}_1\hat{H}\hat{P}=\frac{\Omega_\mathrm{er}^2}{4\Delta}\hat{N}_e, \label{eq_adia_proj_H1}
\end{align}
where $\hat{N}_e=\sum_{i=1}^N \hat{\sigma}_{i}^{ee}$ is the number operator for excited states. Eq.~\eqref{eq_adia_proj_H1} represents the first-order AC Stark shift of a non-interacting atomic ensemble in a far-detuned laser field~\cite{BalewskiNJP14}.

For $n=2$, the Schr\"{o}dinger equation extends up to maximally two Rydberg excitations
\begin{align}
    i\frac{d}{dt}\begin{pmatrix}
        \hat{P}|\psi\rangle\\
        \hat{Q}_1|\psi\rangle\\
        \hat{Q}_2|\psi\rangle\\
    \end{pmatrix}&=\begin{pmatrix}
            \hat{P}\hat{H}\hat{P} & \hat{P}\hat{H}\hat{Q}_1 & \hat{P}\hat{H}\hat{Q}_2 \\
            \hat{Q}_1\hat{H}\hat{P} & \hat{Q}_1\hat{H}\hat{Q}_1 & \hat{Q}_1\hat{H}\hat{Q}_2 \\
            \hat{Q}_2\hat{H}\hat{P} & \hat{Q}_2\hat{H}\hat{Q}_1 & \hat{Q}_2\hat{H}\hat{Q}_2 \\
\end{pmatrix}\begin{pmatrix}
        \hat{P}|\psi\rangle\\
        \hat{Q}_1|\psi\rangle\\
        \hat{Q}_2|\psi\rangle\\
    \end{pmatrix}. \label{eq_sch_Q2}
\end{align}
Applying the quasi-static condition (for $n=1,2$) to Eq.~\eqref{eq_sch_Q2} (similarly to that of Eq.~\eqref{eq_sch_01}) yields \begin{align}
    \nonumber \hat{Q}_1|\psi\rangle&=
    -(\hat{Q}_1\hat{H}\hat{Q}_1)^{-1}
    \big[1+\hat{Q}_1\hat{H}\hat{Q}_2(\hat{Q}_2\hat{H}\hat{Q}_2)^{-1}\\ 
    &\qquad\times\hat{Q}_2\hat{H}\hat{Q}_1(\hat{Q}_1\hat{H}\hat{Q}_1)^{-1}\big]\hat{Q}_1\hat{H}\hat{P}\hat{P}|\psi\rangle, \label{eq_sch_03}
\end{align}
where in the squared bracket, we truncated the Neumann series $\frac{1}{1-\hat{T}}=1+\hat{T}+\hat{T}^2+\dots$ up to the linear order by assuming $\|\hat{T}\|\sim\|(\hat{Q}_{m+1}\hat{H}\hat{Q}_{m+1})^{-1}\|\sim1/(\Delta-mV/2)\ll1$. We note that in general, this assumption for arbitrary orders will impose further limitations $\Delta-mV/2\gg \Omega_\mathrm{er}$ with $m=1,2,\dots$ on the validity of our effective model Eq.~\eqref{eq_Heff_ansatz}, which is discussed in~\cite{supp_mat}. Using Eq.~\eqref{eq_sch_03}, the equation of motion for the excited subspace becomes \begin{align}
    i\frac{d}{dt}\hat{P}|\psi\rangle&=\left(\hat{H}_\mathrm{eff}^{(0)}+\hat{H}_\mathrm{eff}^{(1)}+\hat{H}_{tot,2}\right)\hat{P}|\psi\rangle, \label{eq_sch_02}
\end{align}
where \begin{align}
    \nonumber \hat{H}_{tot,2}&=-\hat{P}\hat{H}\hat{Q}_1(\hat{Q}_1\hat{H}\hat{Q}_1)^{-1}\hat{Q}_1\hat{H}\hat{Q}_2(\hat{Q}_2\hat{H}\hat{Q}_2)^{-1}\hat{Q}_2\hat{H}\hat{Q}_1\\
    \nonumber &\hspace{0.5cm}\times (\hat{Q}_1\hat{H}\hat{Q}_1)^{-1}\hat{Q}_1\hat{H}\hat{P},\\
    &=\frac{\Omega_\mathrm{er}^4}{16\Delta^3(1-\frac{V}{2\Delta})}\hat{N}_e(\hat{N}_e-1). \label{eq_eq_adia_proj_H2_tot}
\end{align}
We can focus only on $\hat{H}_{tot,2}$, as it is the only relevant term consisting of transitions and propagations involving the two-Rydberg $\hat{Q}_2$ subspace; other lower order terms in Eq.~\eqref{eq_sch_02} will cancel each other after taking the difference between the presence $V\neq 0$ and absence $V=0$ of Rydberg interactions for the propagation in the $\hat{Q}_2$ subspace. In the absence of Rydberg interactions $V=0$, $\hat{H}_{\mathrm{tot},2}$ reduces to $\frac{\Omega_{er}^4}{16\Delta^3}\hat N_e(\hat N_e-1)$, which corresponds to the second-order correction to the single-particle light shift of $N_e$ non-interacting atoms, or the second-order AC Stark shift\cite{BalewskiNJP14}. Subtracting this shifting term from Eq.~\eqref{eq_eq_adia_proj_H2_tot} for $V\neq 0$, we effectively renormalize $\hat{H}_{tot,2}$ by the second-order AC Stark sfhit, and extract the pure Rydberg-mediated two-body interaction \begin{align}
    \nonumber \hat{H}_\mathrm{eff}^{(2)}&=\hat{H}_{tot,2}-\hat{H}_{tot,2}\Big|_{(\hat{Q}_2\hat{H}\hat{Q}_2)_0^{-1}}\\
    &=\frac{\Omega_\mathrm{er}^4}{16\Delta^3}\left(\frac{1}{1-\frac{V}{2\Delta}}-1\right)\hat{N}_e(\hat{N}_e-1), \label{eq_adia_proj_H2}
\end{align}
where the subscript $(\hat{Q}_2\hat{H}\hat{Q}_2)_0^{-1}$ denotes the evaluation of $\hat{H}_{tot,2}$ by setting $V=0$ for the propagation in the two-Rydberg subspace. In the Rydberg blockade regime $V\rightarrow\infty$, Eq.~\eqref{eq_adia_proj_H2} becomes $-\frac{\Omega_\mathrm{er}^4}{16\Delta^3}\hat{N}_e(\hat{N}_e-1)$, which corresponds exactly to the Blockade-mediated second-order Kerr nonlinearity that has been studied in~\cite{KeatingPRL16, BalewskiNJP14}.

Repeating this process for higher-order Rydberg subspaces, namely \begin{align}
    \hat{H}_\mathrm{eff}^{(n\geq 2)}&=\hat{H}_{tot,n}-\hat{H}_{tot,n}\Big|_{(\hat{Q}_n\hat{H}\hat{Q}_n)_0^{-1}},
\end{align}
where $\hat{H}_{tot,n}$ is obtained similarly as $\hat{H}_{tot,2}$ from Eq.~\eqref{eq_sch_02}, and contains the transition and propagation in the $\hat{Q}_n$ subspace. Summing up all $n$ body interactions $\hat{H}_\mathrm{eff}^{(n)}$, we obtain the effective Hamiltonian \begin{align}
    \hat{H}_\mathrm{eff}=\sum_{n=2}^N\hat{H}_\mathrm{eff}^{(n)}=\sum_{n=1}^N\mathcal{K}_n\hat{N}_{e}^n\label{eq_adia_proj_Heff_stirling},
\end{align}
where we moved to a rotating frame about the linear term $\mathcal{K}_1\hat{N}_e$, $\mathcal{K}_n=\sum_{m=0}^{N-1}s(m+1,n)f(m)$ is the $n$-th order Kerr nonlinearity, $s(m,n)$ is the (signed) Stirling number of the first kind, and \begin{align}
    f(m)&=\begin{cases}
    \frac{\Omega_\mathrm{er}^2}{4\Delta} &\hspace{-0.27cm}m=0,\\
    \frac{1}{(m-1)!}\frac{V\Omega_\mathrm{er}^{2+2m}}{2^{3+2m}\Delta(\Delta-\frac{m}{2}V)\prod_{k=1}^m(\Delta-\frac{k-1}{2}V)^2}\quad&\text{else}.\hspace*{-0.3cm}
    \end{cases}
\end{align}

\clearpage

\onecolumngrid
\section*{Supplementary Material}
\twocolumngrid

\renewcommand{\theequation}{S.\arabic{equation}}
\renewcommand{\thefigure}{S.\arabic{figure}}
\setcounter{equation}{0}
\setcounter{figure}{0}

\subsection*{Nonlinear evolution analysis}
In this section, we show that the shortest possible time $t_{c,\mathrm{min}}$ for an initial coherent state to evolve into a cat state under multiple orders of Kerr nonlinearities, scales inversely with the factorial $m!$, where $m$ is the highest Kerr order present. We further derive the corresponding configuration of Kerr coefficients that achieves this minimum preparation time.

We consider the Hamiltonian containing multiple orders of Kerr nonlinearities $\mathcal{K}_j$ \begin{align}
    \hat{H}&=\sum_{j=2}^m \mathcal{K}_j (\hat{a}^\dag\hat{a})^j,
\end{align}
where $\hat{a}$ is the bosonic annihilation operator. In Fock basis, an initial coherent state $|\alpha\rangle$ evolves under this Hamiltonian as
\begin{align}
    e^{-i\hat{H}t}|\alpha\rangle&=e^{-\frac{|\alpha|^2}{2}}\sum_{n=0}^\infty \frac{\alpha^n}{\sqrt{n!}}e^{-ip(n)t}|n\rangle, \label{eq_kerr_evol_coherent_pn}
\end{align}
where $p(n)=\sum_{j=2}^m\mathcal{K}_jn^j$ is obtained by applying $\hat{H}$ to the Fock states. Since the cat state \begin{align}
    |\psi_{\mathrm{cat}}\rangle\sim\left(\sum_{\substack{n=\mathrm{even}}}^\infty-i\sum_{\substack{n=\mathrm{odd}}}^\infty\right)\frac{\alpha^n}{\sqrt{n!}}|n\rangle\sim|\alpha\rangle+i|-\alpha\rangle,
\end{align}
exhibits the phase structure $1$ for even $n$ and $-i$ for odd $n$, the condition for the state in Eq.~\eqref{eq_kerr_evol_coherent_pn} to evolve into such a cat state becomes $e^{-ip(n)t_c}=1$ for even $n$ and $e^{-ip(n)t_c}=-i$ for odd $n$, or equivalently
\begin{align}
    p(n)t_c'=\begin{cases}
        0\Mod 2, & n=\mathrm{even},\\
        \frac{1}{2}\Mod 2, & n=\mathrm{odd},
    \end{cases}\label{eq_ivp_mod_condition}
\end{align}
where $t_c'=t_c/\pi$. Here, the notation $r\Mod 2=r+2z$, with $z\in\mathbb{Z}$ and $r\in\{0,\frac{1}{2},1,\frac{3}{2}\}$, accounts for the $2\pi$ periodicity of the phase. The polynomial $p(n)$ can be expressed in the binomial basis, \begin{align}
    p(n)=\sum_{j=2}^m \mathcal{K}_j n^j=\sum_{r=1}^m a_r\binom{n}{r},
\end{align}
where $\mathcal{K}_j$ and $a_r$ are the nonlinear coefficients in the monomial and binomial bases, respectively, which are related by \begin{align}
    \mathcal{K}_j=\sum_{r=1}^m \frac{s(r,j)}{r!}a_r,\label{eq_kerr_coeff_stirling}
\end{align}
where $s(r,j)$ denotes the (signed) Stirling number of the first kind.

We denote $a_rt_c'=\gamma_r$ and insert $n=1,2,...,m$ into $p(n)$ to obtain $m$ equations for $m$ variables $\gamma_{1,2,...,m}$: \begin{align}
    \nonumber \gamma_1&=\frac{1}{2}\Mod 2,\\
    \nonumber 2\gamma_1+\gamma_2&=0\Mod 2,\\
    \nonumber 3\gamma_1+3\gamma_2+\gamma_3&=\frac{1}{2}\Mod 2,\\
    \nonumber &...,\\
    \nonumber \sum_{j=1}^m\gamma_{j}\binom{m}{j}&=\begin{cases}0\Mod 2\hspace{0.5cm}m=\mathrm{even},\\
    \frac{1}{2}\Mod 2\hspace{0.5cm}m=\mathrm{odd}.
    \end{cases}
\end{align}
The solutions to this system of $m$ equations are $\gamma_1=\frac{1}{2}\Mod2$, $\gamma_2=1\Mod2$ and $\gamma_3=\gamma_4=...=\gamma_m=0\Mod 2$, and we can prove by induction that $p(n)$ calculated in this way satisfies Eq.~\eqref{eq_ivp_mod_condition} for all $n$, even though we have only solved for $m$ variables.

Without loss of generality, we set $\mathcal{K}_m=1$ to rescale relevant units such that $\frac{\mathcal{K}_{j<m}}{\mathcal{K}_m}\rightarrow\mathcal{K}_{j<m}$ and time $t\mathcal{K}_m\rightarrow t$ become dimensionless. This allows us to rewrite $\mathcal{K}_mt_c'=1t_c'$ using Eq.~\eqref{eq_kerr_coeff_stirling} as: \begin{align}
    \sum_{r=1}^m\frac{s(r,m)}{r!}\gamma_r&=\frac{\gamma_m}{m!}=t_c'. \label{eq_tc_stirling_sum}
\end{align}
Since $\gamma_r$ are expressed in the modulus form which may have infinitely many possible values, there exists a particular value of $\gamma_r$ for which $t_c'$ in Eq.~\eqref{eq_tc_stirling_sum} is minimized, corresponding to the shortest cat formation time \begin{align}
    t_{c,\mathrm{min}}=\frac{\pi}{\mathcal{K}_m}t_{c,\mathrm{min}}'=\begin{cases}
        \frac{\pi}{2\mathcal{K}_2} &\text{ for $m=2$},\\
        \frac{2\pi}{\mathcal{K}_mm!} &\text{ for $m>2$}.
    \end{cases}
\end{align}
Note that we can also use the same scaling parameter, such as $\mathcal{K}_2$, for all different maximum nonlinear orders. Accordingly, we find \begin{align}
    t_{c,\mathrm{min}}=\frac{\pi}{\mathcal{K}_2}t_{c,\mathrm{min}}'\begin{cases}
        =\frac{\pi}{2\mathcal{K}_2} &\text{ for $m=2,3$},\\
        \propto\frac{2\pi}{\mathrm{lcm}(2,3,...,m)\mathcal{K}_2} &\text{ for $m\geq 4$},
    \end{cases}
\end{align}
where we use $\propto$ to denote an approximate relationship, while a precise formulation would involve further number-theoretic considerations, lying beyond the scope of this discussion.

To determine the required configuration for other $\mathcal{K}_{n<m}$, we observe that Eq.~\eqref{eq_tc_stirling_sum} only imposes a single constraint to $\gamma_m$, implying that we are free to take any values for the other $\gamma_{1,2,...,m-1}$. In other words, we can choose any values for $\gamma_{1,2,...,m-1}$ that satisfy their modulus condition, e.g., $\gamma_3=2,4,6\equiv 0\Mod2$. However, there are always experimental constraints in practice, for example, lower-order nonlinearities can be weaker than higher-order nonlinearities. By solving the system consisting of Eq.~\eqref{eq_tc_stirling_sum} and \begin{align}
    \nonumber |\mathcal{K}_{k< m}|t_c' &\sim |\mathcal{K}_m| t_c'=t_c'\\
    \Rightarrow\left|\sum_{k=1}^m\frac{s(k,n)}{k!}\gamma_k\right|&\sim t_c',
\end{align}
we can obtain possible solutions for $\gamma_{1,2,...,m}$ and hence $\mathcal{K}_j$. Here, $\sim$ indicates possible constraints on lower order nonlinear strengths $\mathcal{K}_{n<m}$, for example, $\sim\ =\ \leq$ indicates $|\mathcal{K}_{k<m}|\leq|\mathcal{K}_m|$ for some $n$.

\subsection*{Validity of the effective Rydberg model}
In this section, we examine the validity of the effective Hamiltonian Eq.~\eqref{eq_adia_proj_Heff_stirling} supported by numerical simulations with realistic experimental parameters~\cite{JohnsonPRA10, LiPRA12, BalewskiNJP14, GlaetzleNC17, SassmannshausenEPJST16, OmranS19}, and thus show that, in the parameter regimes over which it is reliable, strong higher-order Kerr nonlinearities can be obtained and accurately captured by the effective model. To quantify the validity, we evolve the system initialized in the excited subspace $\hat{P}=\prod_{i=1}^N|e_i\rangle\langle e_i|$ under both the full and effective Hamiltonians, and compute their fidelities. An ideal fidelity of unity would indicate that the effective Hamiltonian accurately describes the exact dynamics induced by the full Hamiltonian. More precisely, we consider the initial state
$|\psi_0\rangle=|e_1\rangle\otimes|e_2\rangle\otimes\dots\otimes|e_N\rangle$, and its evolution under the full Hamiltonian Eq.~\eqref{eq_rydberg_full_hamiltonian_Vij} is
\begin{align}
    \nonumber |\psi_\mathrm{full}(t)\rangle&=\hat{P}e^{-i\hat{H}_\mathrm{full}t}|\psi_0\rangle,\\
    \nonumber &=\langle\psi_0|e^{-i\hat{H}_\mathrm{full}t}|\psi_0\rangle|\psi_0\rangle,\\
    &=\sum_k|\langle\psi_0|\phi_k\rangle|^2 e^{-iE_kt}|\psi_0\rangle,
\end{align}
where in the last line, we used eigenstate basis $\{|\phi_k\rangle\}$ of the full Hamiltonian with eigenenergies $E_k$, to make the subsequent numerical simulations more efficient.

Similarly, the evolution of $|\psi_0\rangle$ under the effective Hamiltonian Eq.~\eqref{eq_adia_proj_Heff_stirling} is
\begin{align}
    \nonumber |\psi_\mathrm{eff}(t)\rangle&= e^{-i\hat{H}_\mathrm{eff}t}|\psi_0\rangle,\\
    &=e^{-iE_et}|\psi_0\rangle,
\end{align}
where in the second line, we have used the fact that the fully excited state is already an eigenstate of the effective Hamiltonian, i.e.
\begin{align}
    \hat{H}_\mathrm{eff}|\psi_0\rangle&=\sum_{n=1}^N\mathcal{K}_n\hat{N}_e^n|\psi_0\rangle=\sum_{n=1}^N\mathcal{K}_n N_e^n|\psi_0\rangle=E_e|\psi_0\rangle.
\end{align}

The fidelity between the full and effective Hamiltonian evolutions thus becomes
\begin{align}
    \nonumber \mathcal{F}(t)&=|\langle\psi_\mathrm{eff}(t)|\psi_\mathrm{full}(t)\rangle|^2,\\
    \nonumber &=\left|e^{-iE(N_e)t}\sum_k|\langle\psi_0|\phi_k\rangle|^2 e^{-iE_kt}\right|^2,\\
    \nonumber &=\left|\sum_k|\langle\psi_0|\phi_k\rangle|^2 e^{-iE_kt}\right|^2,\\
    &=\sum_j \delta_j^2|\langle\psi_0|\phi_j\rangle|^4+\sum_{i\neq j}\delta_i\delta_j|\langle\psi_0|\phi_i\rangle|^2|\langle\psi_0|\phi_j\rangle|^2e^{-i(E_i-E_j)t}, \label{Eq_fidelity_comp}
\end{align}
where we note that the summation $\sum_j$ in the last line are over the distinct eigenenergies whose degeneracy is denoted as $\delta_\bullet$. To obtain Eq.~\eqref{Eq_fidelity_comp}, we have also used the fact that the degeneracy arises from permutations of Rydberg states among $N$ atoms, and their overlaps with the excited state subspace remains the same, i.e., $\langle\psi_0|\phi_k\rangle=\langle\psi_0|\phi_{k'}\rangle$, where $|\phi_k\rangle$ and $|\phi_{k'}\rangle$ are two degenerate eigenstates corresponding to the same eigenenergy $E_k$. We observe that the first term in Eq.~\eqref{Eq_fidelity_comp} is the time-averaged fidelity, while the second term corresponds to oscillations induced by the unitary dynamics between different energy levels.

The results are shown in Fig.~\ref{fig_full_eff_comparison}. For small detunings, $0 \leq \Delta/\Omega_\mathrm{er} \leq 8$, the fidelity remains significantly below unity, indicating that the effective model does not accurately reproduce the dynamics of the full system in this regime. This is what we expected, because we have assumed sufficiently large detuning $\Delta/\Omega_\mathrm{er}\gg1$ of the $\Omega_\mathrm{er}$ laser drive to separate the excited state subspace from the single-Rydberg subspace. As the detuning $\Delta/\Omega_\mathrm{er}$ increases, the fidelity gradually approaches unity, demonstrating the asymptotic validity of the effective description in the large-detuning limit required to derive our effective model. Notably, sharp reduction spikes in fidelity are also observed at the discrete detunings $\Delta = mV/2$ for $m=1,2,3,\dots$. These spikes correspond to multi-photon ($m+1$-photon) resonances between the excited-state subspace and the $m+1$-Rydberg subspace. Consequently, they define additional parameter regions in which the effective model fails to accurately capture the full-system dynamics. This breakdown arises because the operators $(\hat{Q}_{n\geq 2}\hat{H}\hat{Q}_{n\geq 2})^{-1}\sim 1/(\Delta-(n-1)V/2)$, which appear in the derivation of the effective Hamiltonian, become singular at these resonances. As a result, the convergence conditions underlying the Neumann-series expansion are violated. This results an overall restriction over the validity of our effective Hamiltonian $\Delta-mV/2\gg \Omega_\mathrm{er}$ for $m=0,1,2,\dots$

However, in the vicinity of multi-photon resonances, the effective Hamiltonian predicts a substantial enhancement of the Kerr nonlinearities. Therefore, our optimal operating regime may be chosen to lie sufficiently close to the resonance conditions, such that the effective model remains quantitatively accurate, while still benefiting from the enhanced Kerr nonlinearities. In practice, one may require the fidelity to remain above a threshold value (e.g., $95\%$). Moreover, as discussed in the previous section, the convergence of the Neumann-series expansion is governed by two competing effects: increasing $\Delta$ suppresses the divergence of the perturbative series, whereas reducing the detuning from a multiphoton resonance, quantified by $|\Delta - mV/2|$, enhances it. Consequently, larger detunings can permit an optimal operation regime closer to the multiphoton resonances, while preserving a high-fidelity effective description, and thereby enabling access to stronger Kerr nonlinearities.

To demonstrate the reliability of the effective Hamiltonian Eq.~\eqref{eq_adia_proj_Heff_stirling} more rigorously, we simulate the Kerr-evolution of an initial (atomic) coherent state \begin{align}
    \nonumber |N,0.5\rangle&\sim(|g\rangle+0.5|e\rangle)^{\otimes N},\\
    &\sim \sum_{n=0}^N \frac{0.5^n}{\sqrt{n!(N-n)!}}\{|e\rangle^{\otimes n}|g\rangle^{\otimes (N-n)}\}_\mathrm{sym},\\
    &\equiv \sum_{n=0}^N \frac{0.5^n}{\sqrt{n!(N-n)!}}|n\rangle,
\end{align}
into a cat state using the full Hamiltonian, and compare it with the same evolution using the effective Hamiltonian. Here, $\{|e\rangle^{\otimes n}|g\rangle^{\otimes (N-n)}\}_\mathrm{sym}$ denotes the symmetrized atomic Dicke state with $n$ atoms in the excited state, and $N-n$ atoms in the ground state~\cite{KeatingPRL16}. We may also denote it simply as the Fock state $|n\rangle$. The results are presented in Fig.~\ref{fig_kerr_cat_evol}(b), where we have worked in an optimal regime in the vicinity of the two-photon resonance spike with a time-averaged fidelity of $98.3\%$. In this regime, the second-order Kerr nonlinearity $\mathcal{K}_2=0.01\mathcal{K}_1$ becomes significant compared to the linear-order term, whereas the third-order Kerr nonlinearity becomes $\mathcal{K}_3=0.01\mathcal{K}_2$. The generated (atomic) cat state can be plotted on a sphere, as well as on a complex plane as the usual bosonic Schr\"odinger's cat state as shown in Fig.~\ref{fig_kerr_cat_evol}(a). This is because the underlying Poisson distribution associated to the bosonic cat state, and binomial distribution for that of the atomic cat state, coincide in the limit of large $N$. In our case, an atomic cat state of amplitude $0.5$ corresponds to a bosonic Schr\"odinger's cat state of $\alpha=\sqrt{2}$.

\begin{figure}
    \includegraphics[width=0.9\linewidth]{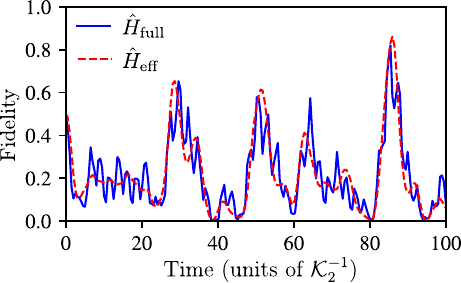}
    \caption{Fidelity evolution of an initial atomic coherent state $|N,0.5\rangle=(|g\rangle+0.5|e\rangle)^{\otimes N}$ with respect to the target cat state $|\psi\rangle\sim (|g\rangle+0.5|e\rangle)^{\otimes N}+i(|g\rangle-0.5|e\rangle)^{\otimes N}$ with $N=10$, under both full and effective Hamiltonians. $\Delta=50.0086\Omega_\mathrm{er}$ and $V=50\Omega_\mathrm{er}$ are used.}
    \label{fig_higher_order_kerr_evol}
\end{figure}

To obtain stronger higher-order Kerr nonlinearities comparable to that of the second order, we need to operate around higher-order multi-photon resonances. For example, we can operate around the three-photon resonance ($\Delta=V$), such that $\Delta=50.0086\Omega_\mathrm{er}$ and $V=50\Omega_\mathrm{er}$ as marked by the second red arrow in Fig.~\ref{fig_full_eff_comparison}. In this case, we obtain $\mathcal{K}_2=0.001\mathcal{K}_1$, together with a strong third-order Kerr $\mathcal{K}_3=-0.526\mathcal{K}_2$. In Fig.~\ref{fig_higher_order_kerr_evol}, we plot the fidelity evolution of an initial atomic coherent state $|N,0.5\rangle$ under both full and effective Hamiltonians, with respect to the target cat state. This is mainly to show that our effective Hamiltonian correctly predicts the presence of strong higher-order Kerr nonlinearities, with a fidelity (or accuracy) of $96.8\%$.

\subsection*{Squeezing analysis}
It has been proposed that applying squeezing to cat states can enhance their robustness against two primary forms of dissipation: single-photon loss and dephasing~\cite{JeannicPRL18, SchlegelPRA22, TehPRR20}. However, the added complexity of managing the combined loss-dephasing dissipative channel has also been noted~\cite{LeviantQ22, MelePRA24}. Here, we propose a simplified approach to facilitate the search for analytical solutions to identify optimal squeezing points.

\begin{figure}[t]
    \includegraphics[width=0.9\columnwidth]{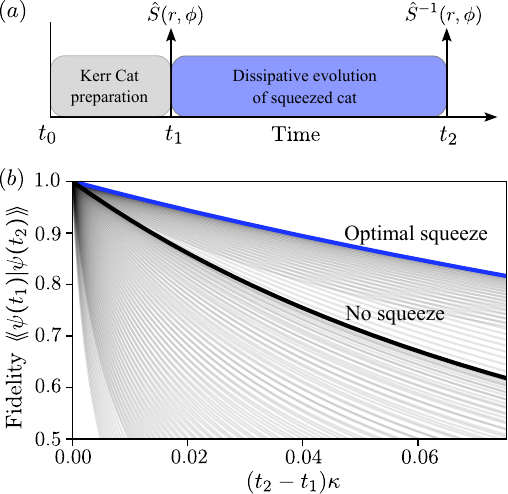}
    \caption[]{(a) Squeezing protocol for protecting cat states. From $t_0$ to $t_1$, a coherent state evolves into a cat state under Kerr nonlinearities, where we assume the preparation time is short enough for us to neglect the influence of dissipation. At $t_1$, the cat state gets squeezed and then evolves under dissipation until $t_2$. At $t_2$, the squeezed cat state, which is no longer a pure state due to dissipation, gets unsqueezed. (b) Fidelity between the pure cat state at $t_1$ and the mixed cat state at $t_2$, against the squeezing time $t_2-t_1$, plotted for different squeezing parameters $0\leq r\leq 1$ and $\phi\in\{0, \text{ }\frac{\pi}{2}, \text{ }\pi\}$. For simplicity, we chose $\alpha=2$ and $\kappa_{1\mathrm{ph}}=\kappa_\phi=\kappa$, and the optimal squeezing occurs at $r\approx 0.51$ and $\phi=0$.}
    \label{fig:opt_squeeze}
\end{figure}

Modeling the photon loss and dephasing using the vectorized Lindblad equation~\cite{DArianoPLA00, AlipourPRL14, QvarfortPRA21}, we can investigate their combined effects on a squeezed cat state using the protocol in Fig.\ref{fig:opt_squeeze}(a). To visualize our discussion, Fig.\ref{fig:opt_squeeze}(b) shows how the fidelity of the cat state decays over the dimensionless time $t\kappa=\kappa_{1\mathrm{ph}}=\kappa_{\phi}$ when being squeezed. The fidelity in Fig.\ref{fig:opt_squeeze}(b) can be approximated with the aid of~\cite{ChaturvediJMO91}, which pointed out that single-photon loss and dephasing channels induce an $\mathfrak{su}(1,1)$ algebra. In other words, we can rewrite the Lindblad equation incorporating single-photon loss and dephasing in terms of the $\mathfrak{su}(1,1)$ algebra generators \begin{align}
    \frac{d}{dt}|\hat{\rho}(t)\rrangle&=\left(-\frac{\kappa_{1\mathrm{ph}}}{2}\hat{K}_3-\frac{\kappa_\phi}{2}\hat{K}_0^2+\kappa_{1\mathrm{ph}}\hat{K}_-\right)|\hat{\rho}(t)\rrangle, \label{eq_lindblad_in_su11_generators}
\end{align}
where $\kappa_{1\mathrm{ph}}$ and $\kappa_\phi$ are single-photon loss and dephasing rates, and \begin{align}
    \nonumber \hat{K}_-&=\hat{a}\otimes\hat{a},\\
    \nonumber \hat{K}_0&=\hat{a}^\dag\hat{a}\otimes\mathds{1}-\mathds{1}\otimes\hat{a}^\dag\hat{a},\\
    \hat{K}_3&=\hat{a}^\dag\hat{a}\otimes\mathds{1}+\mathds{1}\otimes\hat{a}^\dag\hat{a},
\end{align}
are generators of the $\mathfrak{su}(1,1)$ algebra. Here, we have grouped different terms in the Lindblad equation to emphasize its $\mathfrak{su}(1,1)$ structure, which is usually denoted as $\mathfrak{su}(1,1)=\{\hat{K}_0, \hat{K}_3, \hat{K}_+, \hat{K}_-\}$. In our case, $\hat{K}_+=\hat{a}^\dag\otimes\hat{a}^\dag$ corresponds to a thermalized environment. For simplicity, we consider the Lindblad equation under the assumption of a ‘‘cold" environment with no thermal photons $\Bar{n}=0$, but the same approach can be applied to derive solutions for the master equation in cases where $\Bar{n}\neq 0$.

The ansatz for an exact solution to Eq.~\eqref{eq_lindblad_in_su11_generators} is \begin{align}
    |\hat{\rho}(t)\rrangle&=\hat{\mathcal{S}}(t)|\hat{\rho}(0)\rrangle, \label{eq_lindblad_solution_ansatz}
\end{align}
where $\hat{\mathcal{S}}(t)=e^{F_0(t)\hat{K}_0^2}e^{F_3(t)\hat{K}_3}e^{F_-(t)\hat{K}_-}$, and $F_0(t)$, $F_3(t)$ and $F_-(t)$ can be found by evaluating $\left(\frac{d}{dt}\hat{\mathcal{S}}(t)\right)\hat{\mathcal{S}}^{-1}(t)$ in two different ways and then equate them: either inserting Eq.~\eqref{eq_lindblad_solution_ansatz} back to Eq.~\eqref{eq_lindblad_in_su11_generators}, or working directly with the expression of $\hat{\mathcal{S}}(t)$ from Eq.~\eqref{eq_lindblad_solution_ansatz}: \begin{align}
    -\frac{\kappa_{1\mathrm{ph}}}{2}\hat{K}_3+\kappa_{1\mathrm{ph}}\hat{K}_-&=\dot{F}_3(t)\hat{K}_3+\dot{F}_-(t)e^{-2F_3(t)}\hat{K}_-.
\end{align}
Here, since $\hat{K}_0$ is the Casimir element that commutes with all other operators, $F_0(t)$ can be separately calculated \begin{align}
    F_0(t)=-\frac{\kappa_\phi}{2}t.
\end{align}
Equating the coefficients for $\hat{K}_3$ and $\hat{K}_-$, we obtain \begin{align}
    \nonumber F_3(t)&=-\frac{\kappa_{1\mathrm{ph}}}{2}t,\\
    F_-(t)&=1-e^{-\kappa_{1\mathrm{ph}}t}.
\end{align}
Collecting all $F$'s together, we get \begin{align}
    \hat{\mathcal{S}}(t)&=e^{-\frac{\kappa_\phi}{2}\hat{K}_0^2t}e^{-\frac{\kappa_{1\mathrm{ph}}}{2}\hat{K}_3t}e^{(1-e^{-\kappa_{1\mathrm{ph}}t})\hat{K}_-}.
\end{align}

As depicted in Fig. \ref{fig:opt_squeeze}(a), we assess the impact of noise during the period when the cat state is kept in a squeezed state by evaluating the fidelity between the states $|\psi(t_1)\rrangle$ and $|\psi(t_2)\rrangle$: \begin{align}
    \mathcal{F}(t)&=\llangle \psi_\mathrm{cat}|\hat{S}^{-1}(r,\phi)\hat{\mathcal{S}}(t)\hat{S}(r,\phi)|\psi_\mathrm{cat}\rrangle, \label{eq_sq_fid_def}
\end{align}
where $\hat{S}(r,\phi)=e^{\frac{1}{2}r(e^{-i\phi}\hat{a}^{2}-e^{i\phi}\hat{a}^{\dag 2})}$ is the squeezing operator with the squeezing strength $r$ and phase $\phi$.

For small $\kappa_{1\mathrm{ph}}t$ and $\kappa_\phi t$, we can expand Eq.~\eqref{eq_sq_fid_def} up to second order in $\kappa_{1\mathrm{ph},\phi}$: \begin{align}
    \nonumber \mathcal{F}(t)&\approx 1-\kappa_{1\mathrm{ph}}t\langle\hat{N}\rangle-\kappa_\phi t(\langle\hat{N}^2\rangle-\langle\hat{N}\rangle^2)\\
    \nonumber &\hspace{0.2cm}+\frac{\kappa_{1\mathrm{ph}}^2 t^2}{4}\left(\langle\hat{N}^2\rangle+\langle\hat{N}\rangle^2+2|\langle\hat{a}^2\rangle|^2\right)\\
    \nonumber &\hspace{0.2cm}+\frac{\kappa_{\phi}^2 t^2}{4}\left(\langle\hat{N}^4\rangle+3\langle\hat{N}^2\rangle^2-4\langle\hat{N}^3\rangle\langle\hat{N}\rangle\right)\\
    &\hspace{0.2cm}+\frac{\kappa_{1\mathrm{ph}}\kappa_\phi t^2}{2}\left(\langle\hat{N}^3\rangle-\langle\hat{N}\rangle\langle\hat{N}^2\rangle\right), \label{eq_fidelity_expansion_N}
\end{align}
where $\hat{N}=\hat{a}^\dag\hat{a}$ is the number operator, $\langle\hspace{0.1cm}\bullet\hspace{0.1cm}\rangle$ denotes of taking the expectation value with respect to the squeezed cat state, and terms $\langle\hat{a}^{k}\rangle\approx 0$ with odd integers $k$ are ignored. In terms of cumulants, we can rewrite Eq.~\eqref{eq_fidelity_expansion_N} as \begin{align}
    \nonumber \mathcal{F}(t)&\approx 1-\kappa_{1\mathrm{ph}}tC^{(1)}-\kappa_\phi t C^{(2)}\\
    &\hspace{0.5cm}+\frac{\kappa_{1\mathrm{ph}}\kappa_\phi t^2}{2}\left(C^{(3)}+2C^{(1)}C^{(2)}\right)+\mathcal{O}(\kappa_{1\mathrm{ph}}^2,\kappa_\phi^2), \label{eq_sq_fid_exp}
\end{align}
where $C^{n}$ denotes the n-th cumulant of the photon number. In general, any type of non-Hermitian noise $\sqrt{\kappa_{D}}\hat{D}$ contributes a linear term $-\kappa_{D}(\langle\hat{D}^\dag\hat{D}\rangle-\langle\hat{D}\rangle^2)$ to the fidelity.

Here, the terms linear in $\kappa_{1\mathrm{ph}}$ and $\kappa_\phi$ in Eq.~\eqref{eq_sq_fid_exp} indicate that in the presence of only single-photon loss or dephasing, optimal squeezing protects the cat states by minimizing the average photon number $C^{(1)}$ and photon number variance $C^{(2)}$. When both types of dissipation are present, a compromise must be found. This can be done by tuning the squeezing parameter $r$ and its phase $\phi$ to a particular value as calculated in the following. We stress that the choice of $r$ and $\phi$ may depend on the cat state amplitude $|\alpha|$ and the number of cat state components. The cross term $\kappa_{1\mathrm{ph}}\kappa_{\phi}$ also suggests that single-photon loss and dephasing can each amplify the influence of the other, which may relates to the loss-dephasing effect~\cite{LeviantQ22, MelePRA24}.

Now, to concretely derive the optimal squeezing parameter, we first retain only the linear terms in Eq.~\eqref{eq_sq_fid_exp}. The inclusion of higher-order terms proceeds analogously. Without loss of generality, we use the “fractional revival” phenomenon to denote a c-component squeezed cat state~\cite{BosePRA97}: \begin{align}
    |\psi_{\mathrm{SqCat}}^c\rangle&=\hat{S}(r,\phi)e^{\pm i\frac{\pi}{c}(\hat{a}^\dag\hat{a})^2}|\alpha\rangle,
\end{align}
where $\hat{S}(r,\phi)=e^{\frac{1}{2}r(e^{-i\phi}\hat{a}^{2}-e^{i\phi}\hat{a}^{\dag 2})}$ is the squeezing operator, and $\alpha$ is real instead of complex. Then, we find \begin{align}
    \nonumber C^{(1)} &=\langle\hat{a}^\dag\hat{a}\rangle=\langle\psi_{\mathrm{SqCat}}^c|\hat{a}^\dag\hat{a}|\psi_{\mathrm{SqCat}}^c\rangle,\\
    &=\sinh^2(r)+\alpha^2\cosh(2r)-\frac{1}{2}\sinh(2r)\alpha^2A, \label{eq_sq_fid_C1}
\end{align}
and \begin{align}
    \nonumber \langle(\hat{a}^\dag\hat{a})^2\rangle&=\langle\psi_{\mathrm{SqCat}}^c|(\hat{a}^\dag\hat{a})^2|\psi_{\mathrm{SqCat}}^c\rangle,\\
    \nonumber &=\alpha^2(1+\alpha^2)\cosh^2(2r)+\sinh^4(r)\\
    \nonumber &\hspace{0.2cm}+2\alpha^2\sinh^2(r)\left(\cosh(2r)-\frac{1}{2}\sinh(2r)A\right)\\
    &\hspace{0.2cm}-\frac{1}{4}\sinh(4r)\alpha^2B+\frac{1}{4}\sinh^2(2r)D,
\end{align}
where \begin{align}
    \nonumber A &= 2e^{-\alpha^2}\text{Re}[e^{-i\phi}e^{i\frac{4\pi}{c}}e^{\alpha^2e^{i\frac{4\pi}{c}}}],\\
    \nonumber B &= 2e^{-\alpha^2}Re\bigg[e^{-i\phi}e^{i\frac{4\pi}{c}}e^{\alpha^2e^{i\frac{4\pi}{c}}}(\alpha^2+2+\alpha^2e^{i\frac{4\pi}{c}})\bigg],\\
    D &= 2(\alpha^2+1)^2+2\alpha^4e^{-\alpha^2}Re\bigg[e^{-2i\phi}e^{i\frac{16\pi}{c}}e^{\alpha^2e^{i\frac{8\pi}{c}}}\bigg].
\end{align}

The first cumulant $C^{(1)}$ denotes the average photon number of the system, and is therefore just $\langle\hat{a}^\dag\hat{a}\rangle$. The second cumulant $C^{(2)}$ denotes the photon number variance of the system, and is written as \begin{align}
    \nonumber C^{(2)}&=\langle(\hat{a}^\dag\hat{a})^2\rangle-\langle\hat{a}^\dag\hat{a}\rangle^2,\\
    \nonumber &=\alpha^2\cosh^2(2r)-\frac{1}{2}\cosh(2r)\sinh(2r)\alpha^2(B-2\alpha^2A)\\
    &\hspace{0.2cm}+\frac{1}{4}\sinh^2(2r)(D-\alpha^4A^2). \label{eq_sq_fid_C2}
\end{align}
Note that for the 2-component cat state $c=2$, Eq.~\eqref{eq_sq_fid_C1}-\eqref{eq_sq_fid_C2} are identical to that of the squeezed coherent states calculated in~\cite{Alexanian20}.

If we only consider the single photon loss in the absence of dephasing $\kappa_\phi=0$, Eq.~\eqref{eq_sq_fid_exp} indicates that the optimal squeezing for 2-component cat states occurs at the minimum average photon number $C^{(1)}$, so \begin{align}
    \nonumber \frac{\partial}{\partial\phi}C^{(1)}\bigg|_{\phi'}&=0,\\
    \Rightarrow \hspace{0.5cm} \phi'&=0,
\end{align}
and \begin{align}
    \nonumber \frac{\partial}{\partial r}C^{(1)}\bigg|_{r'}&=0,\\
    \Rightarrow \hspace{0.5cm}r'&=\frac{1}{8}\ln(1+8\alpha^2).
\end{align}
Whereas if we only consider dephasing in the absence of single photon loss $\kappa_{1\mathrm{ph}}=0$, Eq.~\eqref{eq_sq_fid_exp} indicates that the optimal squeezing for 2-component cat states occurs at the minimum photon number fluctuation $C^{(2)}$, so \begin{align}
    \nonumber \frac{\partial}{\partial\phi}C^{(2)}\bigg|_{\phi'}&=0,\\
    \Rightarrow \hspace{0.5cm} \phi'&=0,
\end{align}
and \begin{align}
    \nonumber \frac{\partial}{\partial r}C^{(2)}\bigg|_{r'}&=0,\\
    \Rightarrow \hspace{0.5cm}r'&=\frac{1}{4}\ln(1+4\alpha^2).
\end{align}

In the presence of both single photon loss and dephasing, the optimal squeezing conditions can be obtained by minimizing $\kappa_{1\mathrm{ph}}C^{(1)}+\kappa_\phi C^{(2)}$. Here, we ignore the third cumulant $C^{(3)}$ term in Eq.~\eqref{eq_sq_fid_exp} because it corresponds to a higher-order correction via the loss-dephasing channel, and produces only a tiny shift around the optimal squeezing point. Although this channel may play an important role, we choose to ignore it for a simpler illustration of how to calculate the optimal squeezing. We consider the 2-component $c=2$ cat state used in Fig. \ref{fig:opt_squeeze} with the parameters $\kappa_{1\mathrm{ph}}=\kappa_\phi$ and $\alpha=2$, which gives \begin{align}
    \nonumber A&=2\cos(\phi),\\
    \nonumber B&=4(\alpha^2+1)\cos(\phi),\\
    D&=2(\alpha^2+1)^2+2\alpha^4\cos(2\phi).
\end{align}
The optimized squeezing phase $\phi'$ occurs at \begin{align}
    \nonumber \frac{\partial}{\partial\phi}(\kappa_{1\mathrm{ph}}C^{(1)}+\kappa_\phi C^{(2)})\bigg|_{\phi=\phi'}&=0,\\
    \phi'&=0,
\end{align}
for which we can further simplify \begin{align}
    \nonumber A&=2,\\
    \nonumber B&=4(\alpha^2+1),\\
    D&=4\alpha^2(\alpha^2+1)+2.
\end{align}
Thus, we have \begin{align}
    \nonumber \frac{\partial}{\partial r}(\kappa_{1\mathrm{ph}}C^{(1)}+\kappa_\phi C^{(2)})\bigg|_{r=r'}&=0,\\
    1+8\alpha^2+(1+4\alpha^2)e^{2r'}-e^{6r'}-e^{8r'}&=0. \label{eq_sq_fid_opt_para}
\end{align}
While Eq.~\eqref{eq_sq_fid_opt_para} is solvable (quartic in $e^{2r})$, its solution is cumbersome. However, a numerical solution of $r\approx 0.51$ can easily be obtained for our case $\alpha=2$ using mathematical software programs.
\end{document}